\documentclass[lettersize,journal]{IEEEtran}
\usepackage{amsmath,amsfonts}

\usepackage{cite}
\usepackage{array}
\usepackage{algorithm}
\usepackage{algorithmic}
\usepackage{tabularx}
\usepackage{booktabs}

\usepackage{bbding}
\usepackage[caption=false,font=normalsize,labelfont=sf,textfont=sf]{subfig}
\usepackage{multirow}
\usepackage{makecell}
\usepackage{textcomp}
\usepackage{stfloats}
\usepackage{url}
\usepackage{verbatim}
\usepackage{graphicx}
\usepackage[T1]{fontenc} 
\usepackage{amssymb} 
\usepackage{algorithmic}
\usepackage{algorithm}
\usepackage[backref=False]{hyperref}
\def\BibTeX{{\rm B\kern-.05em{\sc i\kern-.025em b}\kern-.08em
    T\kern-.1667em\lower.7ex\hbox{E}\kern-.125emX}}
\usepackage{balance}
\newcommand{\Rmnum}[1]{\uppercase\expandafter{\romannumeral #1}}  
\usepackage{tikz,xcolor,hyperref}

\usepackage{amsmath}
\usepackage{subcaption}
\usepackage{graphicx}
\usepackage{amssymb}

\definecolor{lime}{HTML}{A6CE39}
\DeclareRobustCommand{\orcidicon}{
	\begin{tikzpicture}
		\draw[lime, fill=lime] (0,0)
		circle[radius=0.16]
		node[white]{{\fontfamily{qag}\selectfont \tiny \.{I}D}}; 
	\end{tikzpicture}
	\hspace{-2mm}
}
\foreach \x in {A, ..., Z}{%
	\expandafter\xdef\csname orcid\x\endcsname{\noexpand\href{https://orcid.org/\csname orcidauthor\x\endcsname}{\noexpand\orcidicon}}
}

\begin{document}
\title{Base Station Sleeping Strategy Based on Load Sharing in Ultra-Dense Networks}

\author{Ruixing~Ren\hspace{-1.5mm}\orcidA{},~\IEEEmembership{Graduate~Student~Member,~IEEE}, Shan~Chen, Xuehan~Bao, Pingzheng~Ge, Dongming~Wang,~\IEEEmembership{Member,~IEEE}, Junhui~Zhao,~\IEEEmembership{Senior~Member,~IEEE}

\thanks{
Corresponding author: Junhui Zhao.

Ruixing Ren and Junhui Zhao are with the School of Electronic and Information Engineering, Beijing Jiaotong University, Beijing 100044, China. (e-mail: renruixing0604@163.com; junhuizhao@hotmail.com)

Shan Chen, Xuehan Bao, and Pingzheng Ge are with the School of Information and Software Engineering, East China Jiaotong University, Nanchang 330013, China.

Dongming Wang is with the National Mobile Communications Research Laboratory and also School of Information Science and Engineering, Southeast University, Nanjing 210096, China.

		
}
}

\maketitle

\begin{abstract}
To address the issues of high operational costs and low energy efficiency (EE) caused by the dense deployment of small base stations (s-BSs) in 5G ultra-dense networks (UDNs), this paper first constructs a multi-objective mathematical optimization model targeting maximizing EE and minimizing the number of active BSs. The model incorporates key constraints including BS operational state, user equipment (UE)-BS connection relationship, and load threshold, laying a theoretical foundation for the coordinated optimization of energy conservation and quality of service. Based on this model, an integrated solution combining UE-BS initial connection optimization and load-sharing based BS sleeping is proposed. In the initial connection phase, with communication quality and BS load as dual constraints, efficient matching between UEs and optimal BSs is achieved through three sequential steps: communication feasibility screening, redundant connection removal, and overload load redistribution. This resolves the problems of load imbalance and difficult identification of redundant BSs in UDNs arising from unordered initial connections. In the BS sleeping phase, a BS sleeping index, comprehensively considering UE transferability and backup BS resources, is innovatively introduced to quantify BS dormancy priority. Through a closed-loop process involving low-load BS screening, adjacent BS load evaluation, and load sharing by two takeover BSs based on their capacity, accurate dormancy of redundant BSs and collaborative load migration are realized. Simulation results in a typical UDNs scenario demonstrate that, compared with the traditional baseline scheme, the proposed solution exhibits significant advantages in convergence speed, optimization of the number of active BSs, and EE improvement.
\end{abstract}


\section{Introduction}\label{sec1}
In recent years, the large-scale expansion of the internet of things (IoT) and the rapid evolution of mobile communication technologies have driven global mobile networks into a development phase characterized by high bandwidth and high connection density \cite{IoT}. Currently, mobile networks have undergone a comprehensive upgrade from traditional voice calls and short message services to integrated communication systems supporting high-speed data services, with users’ demands for network capacity and coverage quality continuously escalating \cite{RenIoV}. To address this demand, ultra-dense networks (UDNs) have emerged as a core technology for fifth generation (5G) and beyond communication systems. By deploying small base stations (s-BSs) in high density within a limited area, UDNs significantly enhance network capacity and coverage density while reducing signal fading and transmission latency—achieved through shrinking the coverage radius of BSs and lowering transmission power \cite{srivastava2020energy}. Consequently, UDNs have become a pivotal architecture for supporting massive connections and high-speed data transmission.

However, the ultra-dense deployment of UDNs also brings severe energy and cost challenges. On a global scale, communication network infrastructure already accounts for 3\% of the world’s total energy consumption and contributes 2\% to global carbon dioxide emissions \cite{zagrouba2021comparative}. Among network components, BSs serve as core equipment, with a single BS consuming approximately 4.9 kW of electricity annually. Notably, the radio access segment of mobile communication systems—including BSs—accounts for 70\% of the total energy consumption of network operators, making it the primary source of network energy use \cite{alsamhi2021green}. In UDN scenarios, the long-term operation of a large number of s-BSs under no-load or low-load conditions further exacerbates energy waste and the rise in operational costs, which is seriously contrary to the development concept of green communications that advocates improving energy efficiency (EE) and reducing resource consumption through technical optimization. Therefore, how to reduce the ineffective energy consumption of BSs in UDNs while ensuring network coverage and quality of service (QoS) has become a key issue in promoting the sustainable development of the communication industry.

BS sleeping technology is widely recognized as an effective solution to the aforementioned problems. Its core logic is to dynamically adjust the operational state of BSs (active/sleeping) according to real-time network load, enabling redundant BSs during low-load periods to enter a low-power sleep mode, thereby reducing energy consumption without compromising network performance \cite{10285118}. However, in practical UDN applications, the implementation of this technology faces two core challenges: first, in UDNs, BSs and user equipments (UEs) are densely distributed, and the coverage overlap between BSs is high. If BSs are sleeped blindly, it is likely to cause overload of the remaining active BSs, leading to degradation of QoS. Second, there is a lack of an efficient initial UE-BS association mechanism. If UEs randomly access BSs, it will exacerbate load imbalance, making it difficult to accurately identify truly redundant BSs, which directly affects the optimization effect of sleeping strategies. To address these issues, this paper focuses on the research of BS sleeping strategy based on load sharing in UDN scenarios, aiming to achieve the dual goals of reducing the number of active BSs and improving EE by designing efficient initial association algorithms and sleeping decision mechanisms.

A review of existing studies reveals that BS sleeping technology, as a core approach to reducing energy consumption in cellular networks, has developed multiple technical pathways. However, its applicability in UDN scenarios remains limited. In terms of QoS-balanced sleeping strategies: Wu et al. \cite{wu2020power} proposed an analytical method to quantify the trade-off between power consumption and QoS metrics (e.g., average latency, blocking rate) under different sleeping schemes, providing a theoretical basis for sleeping decisions. Nevertheless, this study failed to consider the load coupling issue caused by the ultra-dense deployment of BSs in UDNs, and its static model cannot adapt to dynamic network changes. Chopra et al. \cite{chopra2020non}, noting that sole reliance on BS sleeping may lead to coverage shrinkage and throughput degradation, proposed introducing non-orthogonal multiple access (NOMA) technology to improve spectral efficiency for load accommodation. However, NOMA is prone to inter-user interference accumulation in UDN scenarios with high user density; furthermore, the study lacked a mechanism for collaborative load sharing among BSs, making it difficult to fundamentally resolve load imbalance and unable to meet the demand for multi-BSs collaborative load bearing in UDNs. Another study by Chopra \cite{Chopra}, focusing on UDNs, proposed a BS sleeping configuration that accounts for UDN deployment scenarios and the number of users per s-BS. Through simulations, it observed the trade-off among downlink rate, fairness index, and energy savings; yet, the work did not elaborate on the design of an inter-BS load sharing mechanism. This limitation results in insufficient improvements in load balance and sleeping decision accuracy in UDN environments characterized by high deployment density and dynamic traffic variations.

With the advancement of artificial intelligence (AI) technology, AI-based intelligent BS sleeping strategies have become a research focus. Such methods capture the dynamic patterns of networks by learning from historical data \cite{ZhuYi,Mao}, aiming to improve the accuracy of sleeping decisions. Palani et al. \cite{palani2020artificial} first optimized the deployment of macro BSs and s-BSs to reduce baseline power consumption, and then proposed an AI-based sleeping algorithm. This algorithm optimizes decisions through learning from historical data to shut down redundant BSs and reduce energy consumption. However, the AI algorithm adopted has limitations in adaptability to complex scenarios and real-time performance of intelligent decisions, and fails to design an inter-BS load sharing mechanism suitable for UDNs. Piovesan et al. \cite{piovesan2020joint} applied algorithms such as imitation learning and Q-learning to BSs powered by renewable energy, realizing the joint optimization of load and energy. Nevertheless, such methods have strong dependence on training data. In UDN scenarios, the model’s prediction accuracy tends to decrease due to sudden load changes and user mobility. Additionally, the lack of an inter-BS load collaboration mechanism may lead to frequent switching between sleeping and awakening modes, which instead increases energy consumption. Sun et al. \cite{10817513} proposed a multi-BSs sleeping strategy integrating adaptive cell zooming, user association, and reconfigurable intelligent surfaces. By modeling with Markov Decision Process and optimizing with multiple AI algorithms, this strategy balances BS energy consumption and latency while significantly reducing energy consumption. However, it also fails to design an inter-BS load sharing mechanism adapted to UDNs, resulting in insufficient capability in dynamic load balancing. Zhu et al. \cite{zhu2021joint} combined traffic prediction with sleeping strategies, optimizing performance by adjusting BS states in advance. However, the randomness of user distribution in UDNs reduces prediction accuracy. Moreover, due to the failure to optimize the initial BS association strategy for users, the problem of load imbalance further amplifies the impact of prediction errors.

Sleeping strategies based on optimization algorithms focus on solving the optimal solution for sleeping decisions through heuristic methods, so as to adapt to the requirements of engineering implementation. Yang et al. \cite{yang2021base} adopted the simulated annealing algorithm to optimize sleep scheduling. Although the algorithm can effectively escape local optimal solutions, it has a slow convergence speed, which makes it difficult to meet the real-time requirements of UDNs. Specifically, when sudden load changes occur, the algorithm cannot adjust the BS status in a timely manner, easily missing the optimal opportunity for load sharing. Post et al. \cite{post2021self} proposed a self-organizing user association and sleeping strategy, which achieves load balancing relying on distributed decision-making on the user side. However, due to the lack of a global load coordination mechanism, in areas with dense BSs in UDNs, it is prone to the situation where users cluster to access some BSs, leading to local load overload and instead exacerbating energy waste. Zhang et al. \cite{zhang2021base} realized BS sleeping based on device-to-device (D2D) cluster communication, reducing the access pressure on BSs by aggregating user traffic. Nevertheless, this scheme is only applicable to user-dense scenarios; in edge areas of UDNs with sparse users, D2D clusters are difficult to form, which cannot create conditions for BS sleeping, resulting in limited applicability of load sharing. Sun et al. \cite{sun2023flexible} improved the uplink EE of fully-decoupled radio access networks through resource collaboration, but failed to consider the signal quality guarantee during user handover. When users switch to adjacent base stations after the original BS enters sleep mode, if the signal-to-interference-plus-noise ratio (SINR) fails to reach the threshold, it will lead to the degradation of QoS, which violates the core goal of load sharing. In addition, Salamatmoghadasi et al. \cite{salamatmoghadasi2024novel} introduced high-altitude platforms as super macro BSs to assist in the load evaluation of sleeping BSs. While this approach improves the accuracy of decisions, it requires additional deployment of high-altitude equipment, which significantly increases network construction costs. As a result, it is difficult to promote on a large scale in UDNs and cannot form a universal load sharing solution.

From the perspective of overall research on EE optimization in UDNs, there exists a notable gap in the integration between existing achievements and BS sleeping technologies. The core shortcomings can be summarized into four key aspects: First, the lack of a load collaboration mechanism. Although ultra-dense small cell networks can enhance capacity and offload 80\% of data traffic from macro BSs through the dense deployment of small s-BSs \cite{zhao2023computation,venkateswararao2019dynamic,safitra2023green}, in-depth analysis of the coupling mechanism among deployment density-load distribution-EE is insufficient. Particularly, there is a lack of specific schemes for the collaborative load sharing among adjacent BSs after some BSs enter sleep mode. The failure to consider the load coupling relationship between BSs leads to the risk of overload on remaining active BSs caused by inappropriate sleep decisions. Second, the insufficient optimization of initial connections. Most studies overlook load scheduling during the initial connection phase; random access of UE directly results in load imbalance and impairs the accuracy of identifying redundant BSs \cite{zhao2023tp}. Even though some research has proposed initial matching connection algorithms \cite{venkateswararao2019dynamic}, they fail to dynamically match UEs with BSs based on communication quality (e.g. SINR) and BS load, thus unable to lay a balanced load foundation for load sharing in subsequent BS sleeping strategies. Third, the limited dynamic adaptability. Existing strategies mostly rely on static models or historical data for optimization, making them incapable of coping with real-time fluctuations in UDNs, such as UE mobility and channel variations \cite{sun2019application}, which easily leads to decision lag in scenarios with sudden load changes. Fourth, the single evaluation dimension. EE optimization is often oriented toward a single objective, such as reducing the number of active BSs or improving the EE of individual BSs. A multi-objective joint optimization framework that integrates reducing active BSs and ensuring load balance has not been established, which tends to trigger contradictions, such as sacrificing load balance for energy conservation or retaining excessive redundant BSs for load balance—making it difficult to achieve optimal overall performance.

To address these issues, this paper proposes an integrated scheme combining UE-BS initial connection optimization and load-sharing based BS sleeping. Through the initial connection algorithm, which takes both SINR and BS load as dual constraints, UEs are matched with the optimal BSs, laying a foundation for balanced load distribution. By quantifying the BS sleeping priority (comprehensively considering UE transferability and backup BS resources), redundant BSs are accurately identified and collaborative load migration is implemented. Ultimately, on the premise of ensuring service quality, the number of active BSs is reduced, the network EE is improved, and the development of UDNs toward a low-energy-consumption and sustainable direction is promoted. Main contributions of this paper:
\begin{itemize}
	\item we adopts dual constraints of qualified communication quality and load balance, and achieves efficient matching between UEs and optimal BSs through three key steps: communication feasibility screening, redundant connection removal, and overload load redistribution. This provides a balanced load distribution foundation for subsequent sleeping strategy, solving the problems of load imbalance and difficult identification of redundant BSs in UDNs caused by unordered initial connections.
	\item By comprehensively considering UE transferability and backup BS resources, this study innovatively introduces a BS sleeping index to quantify the priority of BS dormancy. Meanwhile, through a closed-loop process consisting of low-load BS screening, adjacent BS load evaluation, and load sharing by two takeover BSs based on their capacity, it realizes accurate dormancy of redundant BSs and collaborative load migration while ensuring UE communication quality.
	\item With the joint optimization objectives of maximizing EE and minimizing the number of active BSs, this paper constructs a constrained mathematical optimization model combined with the characteristics of UDNs. The model includes constraints such as BS status, UE-BS connection, and load threshold, realizing the coordinated optimization of reducing energy consumption and ensuring QoS, which makes up for the defect of overall performance imbalance in existing studies caused by single-objective optimization.
	\item In a typical UDNs scenario, compared with the traditional baseline scheme, the proposed scheme shows significant advantages in convergence speed, optimization degree of the number of active BSs, and network EE. This fully verifies the practicality and stability of the scheme in dynamic network environments, providing a verifiable technical path and performance support for the green deployment of UDNs.
\end{itemize}

The structure of this paper is as follows. Section 2 introduces the system model. Section 3 conducts a detailed analysis of the UE-BS initial connection algorithm and the load-sharing based BS sleeping algorithm. Section 4 presents the simulation results and analysis. Section 5 is a summary and includes a discussion on future work.

\begin{figure*}[htb]
	\centering     
	\vspace{0cm}
	\hspace*{0cm}
	\setlength{\abovecaptionskip}{0cm} 
	\setlength{\abovecaptionskip}{0cm} 
	\includegraphics[width=0.75\textwidth]{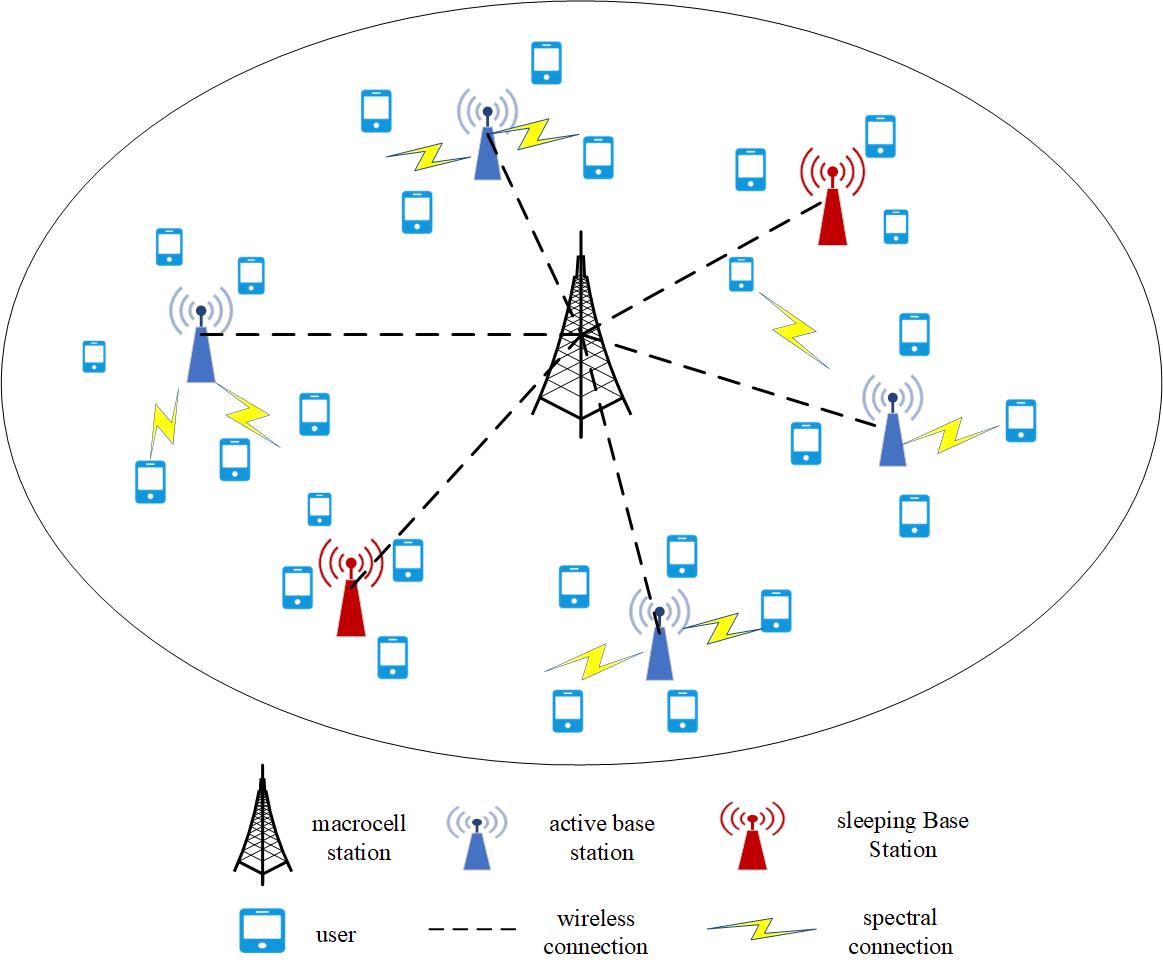}
	\caption{Ultra-Dense Cellular Network System Model.}
	\label{fig1}
\end{figure*}
\section{System model} 
In this paper, an UDNs region is considered, and the network model of this cellular network is illustrated in Figure 1. In this model, multiple micro cells and UEs are managed by a single macro cell \cite{RenRIS}. The set of micro cells in this network is denoted as $\mathbf{N}$ = $\{n_i\}$, $\forall$ $i$ = 1, $\ldots$, $N$. Similarly, the set of UEs is denoted as $\mathbf{M}$ = $\{m_j\}$, $\forall$ $j$ = 1, $\ldots$, $M$. The state of a BS is represented by a matrix $\mathbf{S} = \left\lbrack s_{i} \right\rbrack_{1\times N} $. If $s_{i} = 1$, it indicates that the $i$-th BS is in an active state; otherwise $s_{i} = 0 $ implies that the $i$-th BS is in a sleep state. The connection relationship between the $i$-th BS and the $j$-th UE is represented by matrix $\mathbf{A} = \left\lbrack a_{i,j} \right\rbrack_{M \times N} $. If $ a_{i,j} = 1 $, it means that the $j$-th UE is in a connected state with the $i$-th BS; otherwise, $a_{i,j} = 0$ indicates that there is no connection between the $i$-th BS and the $j$-th UE.

The SINR between the $ i $-th BS and the $ j $-th UE in the network model can be expressed as
\begin{equation}
	\operatorname{SINR}_{i,j} = \frac{a_{n,m} \, s_n \, p_{i,j} \, g_{i,j}}{\displaystyle\sum_{i=1, i \neq n}^{N} \displaystyle\sum_{j=1, j \neq m}^{M} s_i \, p_{i,k} \, g_{i,j} + \sigma^2},
\end{equation}
where $ p_{i,j}^{~}  $ denotes the transmission power assigned by the $ i $-th BS for $ j $-th UE, Coefficient ${a_{n,m}}$ is a parameter associated with the receiving antenna and channel characteristics, typically encompassing factors such as antenna gain and transmission path loss. The transmission power of sender $s_{n}$ is denoted as $n$. $ g_{i,j}^{~} $ denotes the channel gain between the $ i $-th BS and the $ j $-th UE, and $ \sigma^2 $ denotes the Gaussian white noise power.

When the SINR value between the $ j $-th UE and the $ i $-th BS exceeds the threshold $ \theta_{th} $, UE is linked to BS. The equation $R_{i,j} = \log_2 \left( 1 + \text{SINR}_{i,j} \right) $ represents the Shannon capacity that UE achieved at this particular moment \cite{zhao2019computation}. In addition, Equations (2) represents the EE of cellular networks.
\begin{equation}
	E E = \frac{R_{total}}{P_{total}}.
\end{equation}

The throughput $R_{total}$ is the total transmission rate that the entire cellular network system can achieve during a certain period of time. This includes all the data transmitted by connected UEs, devices, or BSs in that time period. The power consumption $R_{total}$ is the total power consumed by the entire network system during the same time period. The higher the EE, the less energy consumed per unit transmission rate, making the system more energy-efficient in energy utilization.

The total data transmission rate is represented by Equation (3), which is the sum of the transmission rates provided by all BSs to all UEs:
\begin{equation}
	R_{total} = \sum_{i=1}^{N} \sum_{j=1}^{M} s_i a_{i,j} R_{i,j},
\end{equation}
where $R_{i,j}$ represents the transmission rate between UE $j$ and BS $i$. The total power consumption is represented by Equation (4), which covers the power consumption of all BSs, including the transmission power and static power in active state, as well as the power consumption in sleep state:
\begin{equation}
	P_{total} = \sum_{i=1}^{N} \left[ s_{i} (p_{i} + p_{i}^{a}) + (1-s_{i}) p_{i}^{s} \right],
\end{equation}
where $ p_{i} $ is the total energy consumption of the $ i $-th BS in the network, and $ p_{i}^{a}/p_{i}^{s} $ denotes the power consumption of the circuits of the BS when it is in the active mode versus the sleep mode, respectively. Equation (5) represents the transmission power of BS. The transmission power $p_i$ of the $i$-th BS is a function of its maximum transmission power $p_i^{\text{max}}$ and the load, where the load is determined by the number of connected UEs. It is given by:
\begin{equation}
	p_{i}{=}{p}_{{i}}^{{max}}{\sum\limits_{{j=1}}^{{M}}\mspace{2mu}}\frac{{a}_{i,j}^{\ }}{{S}_{{max}}},
\end{equation}
where $ p_{i}^{max} $ denotes the maximum transmission power of the $i$-th BS, and $S_{max}$ represents the maximum number of UEs that a single BS can serve simultaneously.

To minimize the communication cost and energy consumption of UDNs without compromising UEs' transmission rate requirements, BS sleep strategy serves as a core technical measure. The essence of this strategy lies in dynamically adjusting the operational status of BSs (active or sleeping) to reduce ineffective energy consumption while ensuring service quality, and this optimization logic can be realized through the mathematical modeling of two key objectives:
\begin{itemize}
	\item Firstly, EE, defined as the ratio of total transmission rate to total power consumption (see Equation (2)), is a core indicator for measuring the energy-saving effect of the network. Maximizing EE can directly quantify the improvement of energy utilization efficiency by the BS sleep strategy, ensuring that sleep operations reduce energy consumption without sacrificing communication performance.
	\item Secondly, the number of active BSs, $\sum_{i = 1}^{N} s_i$, is positively correlated with network energy consumption, reducing the number of active BSs can directly lower overall power consumption, which is precisely the core operational goal of the BS sleep strategy. Minimizing the number of active BSs can accurately reflect the effect of the sleep strategy in shutting down redundant BSs.
\end{itemize}

Therefore, Equations (6) and (7) transform the core objectives of the BS sleep strategy into quantifiable and solvable mathematical optimization problems from the two dimensions of energy utilization efficiency and resource configuration optimization, jointly forming the mathematical expression basis of this strategy, as follow.
\begin{equation}
	{Maximize}\ O_{1}, \mathrm{where} \; O_1= EE,
\end{equation}
\begin{equation}
	{Minimize~}O_{2}, \mathrm{where} \; O_2=\sum_{i = 1}^{N}s_{i}.
\end{equation}

Based on the above equations, the optimization problem can be expressed as:
\begin{equation}
	\begin{aligned} 
		&F_{1} = \left\lbrack {{Maximize}\ O_{1} ,{Minimize~}O_{2}} \right\rbrack\\
		s.t. \ &C1: s_{i}^{~} \in \left\{ 0,1 \right\},\forall i = 1, \ldots , N\\
		&{C2:a_{i,j} \in \left\{ 0,1 \right\},\forall i = 1, \ldots , N;\forall j = 1, \ldots , M}\\
		&{C3:\sum_{i = 1}^{N}a_{i,j} = 1,\forall j = 1, \ldots , M}  \\
		&{C4:\sum_{j = 1}^{M}a_{i,j} \leq s_i \cdot S_{max}, \forall i = 1, \ldots , N} \\
		&{C5:\sum_{i = 1}^{N}\left( {\left( {\sum_{i = 1}^{M}a_{i,j}} \right)/S_{max}} \right) < T_{load}}
	\end{aligned}
\end{equation}
where constraint $C1$ denotes the active/sleep state of the BS, and $C2$ denotes whether there is a connection between the $i$-th BS and the $j$-th UE. Constraint $C3$ denotes that each UE can only establish a connection with one BS. The meaning of $C4$ is that the number of UEs connected to the $i$-th BS shall not exceed the product of its maximum capacity $S_{max}$ and its own active state $s_i$. Specifically, the number of connected UEs of a dormant BS ($s_i = 0$) is forced to be 0, while the number of connected UEs of an active BS ($s_i = 1$) shall not exceed its maximum capacity $S_{max}$. Constraint $C5$ denotes that the sum of the load ratios of all BSs, where the load ratio of each BS is the number of connected UEs divided by its maximum capacity $S_{max}$, is less than a predetermined maximum total load threshold $T_{load}$.

\section{Algorithm Design and Implementation} 
To achieve the core objectives of optimizing EE and reducing the number of active BSs in UDNs, two key issues need to be addressed: optimization of initial UE-BS connections and BS sleeping strategy. This section designs algorithms for these two issues respectively, clarifying the design logic, technical details, and execution flow of each algorithm to ensure that under the premise of guaranteeing user service quality, network energy consumption is reduced and resource utilization is improved.

\subsection{UE-BS Initial Connection Algorithm}

\begin{algorithm}
	\caption{UE-BS Initial Connection Algorithm}
	\renewcommand{\algorithmicrequire}{ \textbf{Input:}}
	\renewcommand{\algorithmicensure}{ \textbf{Output:}}
	\begin{algorithmic}[1] 
		\REQUIRE UE set $\mathbf{M}$, BS set $\mathbf{N}$, SINR threshold $\theta_{th}$, maximum BS capacity $S_{max}$.
		\ENSURE UE-BS connection network (represented by connection matrix $\mathbf{A}$).\\
		\textbf{Initialize:} UEs node degree $\sum_{j = 1}^{M} a_{i,j}$, BSs node degree $\sum_{i = 1}^{N} a_{i,j}$, number of remaining subscribers, subscriber-BS connection network.
		\WHILE{The UE-BS connection network is empty}
		\FOR{$i=1$ to $N$}
		\FOR{$j=1$ to $M$}
		\STATE Calculate the $\mathrm{SINR}_{i,j}$ between the $j$-th UE and $i$-th BS using Equation (1);
		\IF{$\mathrm{SINR}_{i,j}$ > $\theta_{th}$}
		\STATE Establish connection: $a_{i,j} = 1$;
		\STATE Update UE node degree: $\sum_{j = 1}^{M} a_{i,j}=\sum_{j = 1}^{M} a_{i,j} + 1$;
		\STATE Update BS node degree: $\sum_{i = 1}^{N} a_{i,j} = \sum_{i = 1}^{N} a_{i,j} + 1$;
		\ENDIF
		\ENDFOR
		\ENDFOR
		\FOR{$i=1$ to $N$}
		\FOR{$j=1$ to $M$}
		\IF{$a_{i,j} = 1$ and $\sum_{j = 1}^{M} a_{i,j} > 1$}
		\STATE Find the BS $i_{max}$ with highest SINR among those connected to the $j$-th UE;
		\STATE Keep only the connection with $i_{max}$ and disconnect from other BSs;
		\STATE Update UE-BS node degree;
		\ELSE
		\STATE Add $a_{i,j}$ to connected network;
		\ENDIF
		\ENDFOR
		\IF{$\sum_{i = 1}^{N} a_{i,j} > S_{max}$}
		\STATE Sort the UEs in ascending order according to SINR; 
		\STATE Retain only the last $S_{max}$ UEs in the sorted list; 
		\STATE Reallocate the remaining UEs to other BSs with remaining capacity; 
		\STATE Update the node degrees of relevant BSs and $\mathbf{A}$.
		\ENDIF
		\ENDFOR
		\ENDWHILE
	\end{algorithmic}
\end{algorithm}

The initial connection serves as the foundation for the BS sleep strategy: if UEs connect to BSs randomly, it will easily lead to overload in some BSs (exceeding the maximum capacity $S_{max}$) and underload in others (with idle resources). This not only impairs UE communication quality but also makes it difficult for subsequent sleep decisions to identify truly redundant BSs. Therefore, it is necessary to design an initial connection algorithm. By quantifying the communication feasibility between UEs and BSs as well as the load status of BSs, an initial connection network with balanced load and qualified communication quality is established, which provides a reasonable network topology basis for subsequent load sharing and BS sleep operations.

The proposed algorithm takes satisfactory SINR as the communication quality constraint and BS load within the limit as the resource utilization constraint. It achieves the optimal initial connection through the following logic:
\begin{itemize}
	\item Communication Feasibility Screening: Calculate the SINR between each UE and each BS using Equation (1). Only BSs with SINR exceeding the threshold $\theta_{th}$ are retained as candidate connection objects to ensure that the communication quality after connection meets the requirements.
	\item Load Balance Control: Count the number of connected UEs (node degree) of each candidate BS in real time. If the number of connections of a BS reaches its maximum capacity $S_{max}$, the BS will stop accepting new UEs to avoid overload \cite{gures2022machine}.
	\item Redundant Connection Clearing: If a single UE is connected to multiple BSs due to coverage overlap, the connection between the UE and the BS with the highest SINR is retained, while other redundant connections are disconnected to avoid resource waste.
\end{itemize}

To implement the above-designed logic, Algorithm 1 achieves efficient initial connections between UEs and BSs through a phased execution process. The specific operations revolve around three core stages: communication feasibility screening (Lines 3–12), redundant connection clearing (Lines 14–22), and overload control for BS load (Lines 23–28). Specifically, all BSs and UEs are first traversed, and the SINR value between each UE-BS pair is calculated using Equation (1). Only when the SINR exceeds the threshold $\theta_{th}$ is a connection marked in the connection matrix (i.e., $a_{ij} = 1$), with simultaneous updates to the UE node degree (total number of BSs connected to a single UE) and the BS node degree (total number of UEs connected to a single BS). This step ensures that all initial connections meet the minimum requirement for communication quality, preventing service degradation caused by weak signals.

Next, aiming at the one UE connected to multiple BSs phenomenon caused by coverage overlap in UDNs, we performs a secondary verification on connected users: if the node degree of a UE is greater than 1 (i.e., connected to multiple BSs), only the connection with the BS having the highest SINR is retained, while links with other BSs are disconnected, and the relevant node degrees are updated. This operation not only eliminates resource waste but also ensures that each UE can obtain optimal communication quality.

Finally, the node degrees of BSs are monitored in real time. If the number of connected UEs of a BS exceeds its maximum capacity $S_{max}$, all UEs under this BS are sorted in ascending order of their SINR values. Only the connections of the $S_{max}$ UEs with the highest SINR are retained, while all other connections are disconnected and the network status is updated. This step strictly limits the upper limit of BS load, preventing service interruptions caused by overload.

\subsection{Load-Sharing Based BS Sleeping Algorithm}
After establishing the initial connection network between UEs and BSs through the UE-BS initial connection algorithm in Section 4.1, it is necessary to further optimize network energy consumption through BS sleeping strategy. This section designs a load sharing-based BS sleeping algorithm (see Algorithm 2), whose core idea is as follows: redundant BSs are screened by quantifying the sleep feasibility of BSs, and then the redundant BSs are put into sleep mode through load sharing among neighboring BSs. Ultimately, the number of active BSs is reduced and network energy consumption is lowered on the premise of ensuring UE service quality.

To accurately determine whether a BS meets the conditions for sleep, a BS sleep index $\beta_i$ is introduced to quantify the sleep priority of the $i$-th BS. This index comprehensively considers two key factors: UE transferability of the BS and backup BS resources. Its calculation formula is as follows:
\begin{equation}
	{\beta _i} = {k_i} \times {l_i},
\end{equation}
where the UE transferability parameter $k_i$ takes a value of 0 or 1, which is used to determine whether all UEs connected to the $i$-th BS can be switched to other backup BSs. If $k_i = 1$, it indicates that all UEs of the BS can find available backup BSs, satisfying the basic conditions for sleep; if $k_i = 0$, it means there exist UEs that cannot be transferred, and the BS is not eligible for sleep. The backup BS quantity parameter $l_i$ represents the total number of available backup BSs for all UEs connected to the $i$-th BS. The backup BSs herein must meet two conditions: 1) being in an active state; 2) having a SINR with the UE exceeding the threshold $\theta_{th}$, so as to ensure that the communication quality meets the standard after UE handover. A larger $\beta_i$ indicates a higher sleep priority of the $i$-th BS. On one hand, $k_i = 1$ ensures that UE services remain uninterrupted after the BS enters sleep mode; on the other hand, a larger $l_i$ means more alternative backup BSs for UEs, leading to stronger flexibility in load sharing and less impact of the BS's sleep state on the network.

\begin{algorithm}
	\caption{Load Sharing-Based BS Sleeping Algorithm}
	\renewcommand{\algorithmicrequire}{ \textbf{Input:}}
	\renewcommand{\algorithmicensure}{ \textbf{Output:}}
	\begin{algorithmic}[1] 
		\REQUIRE UE set $\mathbf{M}$, BS set $\mathbf{N}$, Initial connection matrix $\mathbf{A}$, $S_{max}$, $\theta_{th}$.
		\ENSURE Final active BS set.\\
		$\mathbf{Initialize:}$ active BS set, BS load degree ${load}_{bs}$, takeover BS $
		BS_{a}^{pair}$ and $BS_{b}^{pair}$, minimum load degree BS $BS^{\text{least~}}$, set of neighboring BS.
		\WHILE{active BS set is not empty}
		\FOR{$i=1$ to $N$}
		\STATE Calculate $load_{bs}[i]$ by Equation (10);
		\STATE Calculate $\beta_i$ by Equation (9);
		\ENDFOR
		\STATE Select sleep candidate BS: $BS^{max}$=argmax($\beta_i$, $i$ $\in$ active BS set);
		\STATE Local search within the transmission radius of $BS_{max}$, and add all other active BSs found to the neighboring BS set;
		\IF{|neighboring BS set| < 1}
		\STATE break
		\ENDIF
		\STATE Sort neighboring BS set in ascending order by $load_{bs}$, select the BS with the smallest load as $BS^{least}$, and record $load_{min} = load_{bs}[BS^{least}]$;
		\STATE Local search within the transmission radius of $BS^{least}$, add all other active BSs found to surrounding BS set;
		\IF{|surrounding BS set| < 2}
		\STATE break
		\ENDIF
		\STATE Sort surrounding BS set in ascending order by $load_{bs}$, select the top two BSs as $BS^{pair}_a$ and $BS^{pair}_{b}$;
		\STATE Calculate traffic sharing ratios for takeover BSs by Equations (11,12);
		\IF{$load_{bs}[BS^{pair}_{a/b}]+L_{a/b}\le 1$, and $\forall j \in \mathbf{M}$,  $\max{(\mathrm{SINR}(BS^{pair}_a,j),\mathrm{SINR}(BS^{pair}_b,j))} \ge \theta_{th}$}
		\STATE Switch $BS^{least}$ to sleep state and remove it from active BS set;
		\STATE Disconnect links between $BS^{least}$ and its originally connected UEs;
		\STATE Establish new connections between each UE and the corresponding takeover BS based on $L_a$ and $L_b$;
		\STATE $\mathbf{continue}$
		\ELSE
		\STATE Sort active BS set in ascending order by $load_{bs}$;
		\IF{|active BS set|>1}
		\STATE select the BS with the second smallest load;
		\ELSE
		\STATE break.
		\ENDIF
		\ENDIF
		\ENDWHILE
	\end{algorithmic}
\end{algorithm}

Algorithm 2 takes the initial connection network between UEs and BSs, the UE set $\mathbf{M}$, and the BS set $\mathbf{N}$ as inputs, and outputs the final active BS set. The overall process proceeds in an orderly manner around four core stages: "screening active BSs, identifying sleep candidate BSs, configuring load sharing, executing sleep decisions". First, initialization and active BS screening are performed. Key parameters such as the active BS set, BS load degree $load_{bs}$, BS with the minimum load degree $BS^{least}$, neighboring BS set, and BS sleep index $\beta_i$ are initialized first. Then, all BSs, $i=1,2,\dots,N$, are traversed: if the $i$-th BS is in the active state, $s_i=1$, its load degree is calculated in accordance with Equation (10), as follows:
\begin{equation}
	{load}_{bs} = \frac{\sum_{j = 1}^{M}a_{i,j}}{S_{max}}.
\end{equation}
This equation quantifies the current load level of the BS using the ratio of the sum of the $i$-th row of the connection matrix $\mathbf{A}$ (i.e., the number of UEs connected to the $i$-th BS) to the maximum UE capacity $S_{max}$ of the BS. The load degree ranges from 0 to 1, where a smaller value indicates a higher redundancy of the BS. Subsequently, this active BS is added to the active BS set.

Next, the process proceeds to the stage of screening high-priority sleep candidate BSs. First, for each BS in the active BS set, the sleep index $\beta_i$ is calculated using Equation (9). Then, the active BSs are sorted in descending order of $\beta_i$, and the top-ranked BS is selected as the high-priority sleep candidate BS $BS^{max}$ with its corresponding correlation coefficient $\beta_i^{max}$ determined. A local search is then performed within the transmission radius of this BS, and all other BSs found in the search are included in the neighboring BS set.

The BSs in the neighboring BS set are sorted in ascending order of load degree in accordance with Equation (10), and the BS with the minimum load degree is selected and denoted as $BS^{least}$, with its load degree recorded as $load_{min}$. The reason for selecting $BS^{least}$ rather than $BS^{max}$ as the sleep target is to avoid load imbalance while ensuring energy-saving efficiency. $BS^{least}$ features the lowest resource utilization and light traffic load, so its sleep has minimal impact on throughput and it can be easily taken over by other BSs; in contrast, if $BS^{max}$ is put into sleep directly, its heavy traffic load is likely to cause overload of the takeover BSs.

Subsequently, a local search is conducted within the transmission radius of $BS^{least}$, and all other BSs identified in the search are incorporated into the surrounding BS set. The BSs in this surrounding BS set are then sorted again in ascending order of load degree, and the top two BSs from the sorted results are selected as takeover BSs, denoted as $BS_a^{pair}$ and $BS_b^{pair}$. Prioritizing the selection of low-load BSs helps avoid overload after takeover. To achieve balanced traffic allocation, the traffic load of $BS^{least}$ needs to be shared proportionally between the two takeover BSs based on their load-bearing capacities. The core design logic is that a stronger load-bearing capacity (i.e., a lower current load degree) corresponds to a larger share of the traffic load. Therefore, the load degree of the counterpart takeover BS is used as the numerator to calculate the respective traffic sharing ratios. The specific formulas for the traffic share ratios $L_a$ and $L_b$ to be borne by the two takeover BSs are as follows:
\begin{equation}
	\begin{split}
		L_{a} = \frac{{load}_{b}^{pair}}{{load}_{a}^{pair} + {load}_{b}^{pair}}{load}_{min} 
	\end{split}
\end{equation}
\begin{equation}
	\begin{split}
		L_{b} = \frac{{load}_{a}^{pair}}{{load}_{a}^{pair} + {load}_{b}^{pair}}{load}_{min}  
	\end{split}
\end{equation}
where $load_a^{pair}$ and $load_b^{pair}$ represent the load degrees of $BS_a^{pair}$ and $BS_b^{pair}$ respectively. For example, if the load degree of $BS_a^{pair}$ is $load_a^{pair} = 0.2$ (with 80\% remaining capacity) and that of $BS_b^{pair}$ is $load_b^{pair} = 0.3$ (with 70\% remaining capacity), while the load degree of $BS^{least}$ is $load_{min} = 0.5$, the calculated results via the formulas will be $L_a = 0.3$ and $L_b = 0.2$. This means $BS_a^{pair}$, with stronger load-bearing capacity, undertakes more traffic load, which conforms to the load balancing principle and ensures that the load degrees of both BSs remain within a reasonable range (not exceeding 1) after load sharing.

Finally, the sleep decision and network state update are executed. If the two takeover BSs successfully take over all traffic from $BS^{least}$, that is, the load degrees of both BSs after takeover are $\le 1$, and the SINR between UEs and the takeover BSs after handover is $\ge$ $\theta_{th}$, then $BS^{least}$ is switched to the sleep state ($s_i = 0$), removed from the active BS set, and the connection matrix $\mathbf{A}$ is updated synchronously (disconnecting the links between $BS^{least}$ and its originally connected UEs, and establishing new connections between each UE and the corresponding takeover BS). If the load sharing fails (including scenarios such as overload of takeover BSs, no available active takeover BSs in the vicinity, or unsatisfactory SINR of UEs after handover), the sleep operation for $BS^{least}$ is abandoned. The current active BS set is re-sorted in ascending order of load degree, and the second-ranked BS (with the second smallest load degree) is selected as the new $BS^{least}$. The complete process of "determining the new $BS^{least}$ $\rightarrow $ searching for neighboring active BSs $\rightarrow $ screening takeover BSs and sharing traffic $\rightarrow $ making sleep decisions" is repeated until, after sorting the current active BS set in ascending order of load degree, there is no $BS^{least}$ that can be successfully put into sleep. In other words, the BS with the smallest load degree among all BSs cannot find a takeover object that meets QoS requirements, and the algorithm terminates.

\section{Simulation results and analysis}
\subsection{Simulation Scenario and Parameter Design}
To verify the effectiveness of the proposed load sharing-based BS sleeping algorithm, a simulation platform is built based on a typical UDNs scenario. A random user distribution model is adopted, which conforms to the actual characteristic of non-uniform user distribution in urban areas. The core parameter configurations are listed in Table 1.
\begin{table*}[htbp]
	\centering
	\caption{Simulation Parameter Configuration}
	\label{tab:simulation_params}
	\begin{tabular}{|l|l|l|}
		\hline
		\textbf{Parameter Name} & \textbf{Value} & \textbf{Description} \\
		\hline
		Simulation Area Size & $1000 \times 1000\ \text{m}^2$ & Square area, simulating urban microcellular dense coverage \\
		\hline
		Number of Microcells & 20 & Uniform deployment, no coverage blind spots \\
		\hline
		BS Communication Radius & 120 m & Consistent with micro BSs' short-coverage and low-power features \\
		\hline
		Maximum User Load per BS ($S_{\text{max}}$) & 30 per BS & Refer to the micro BS capacity standard in 3GPP protocol \\
		\hline
		User Distribution Model & Random Distribution & Aligning with the user distribution patterns in the real world \\
		\hline
		SINR Threshold ($\theta_{th}$) & -5 dB & Ensure minimum UE communication rate \\
		\hline
		Path Loss Exponent & 2 & Typical value for urban dense building environment \\
		\hline
		Noise Power & -174 dBm/Hz & Thermal noise floor value \\
		\hline
		BS Active Mode Power Consumption & 25 W & Including signal transmission, baseband processing, etc. \\
		\hline
		BS Sleep Mode Power Consumption & 8 W & Only retain the standby power consumption of core circuits \\
		\hline
		Channel Gain  & 1  & Simplifying the channel attenuation modeling in simulation  \\
		\hline
	\end{tabular}
\end{table*}

Based on the system model in Section 3 (Equations (1)-(8)), a joint link-level and system-level simulation is implemented. Three core indicators are selected to verify the algorithm performance: (1) Number of active BSs: reflecting the redundancy of network resources, the fewer the number, the better the energy-saving effect; (2) Network EE: calculated according to Equation (2), reflecting the service carrying capacity per unit energy consumption; (3) Number of iterations for algorithm convergence: measuring the optimization efficiency, the fewer the number, the stronger the real-time performance of the algorithm, and the more suitable for dynamic network scenarios.

Selection of Baseline Algorithm: A typical random BS sleeping algorithm from existing studies \cite{xu2016toward} is chosen for comparison. This algorithm randomly selects BSs to sleep, i.e., randomly deactivates a BS and then re-establishes connections with UEs. However, this strategy has a significant drawback: there is no restriction on the frequency at which UEs select BSs during reconnection, which forces the system to perform multiple rounds of operations to find suitable offloading BSs for UEs. This not only increases the complexity of the algorithm but also prolongs the optimization process, thereby affecting user experience and system performance.

\subsection{Simulation Results and Key Analysis}
Figures 2–4 present network topology snapshots at different stages, intuitively demonstrating the optimization effect of the proposed algorithm on network structure. Figure 2 shows the initial deployment state, where all 20 green triangles (representing active small BSs, Active s-BS) are activated, and blue dots (representing UEs) are randomly distributed within the simulation area. It can be observed from the topology that some BSs suffer from load imbalance due to coverage overlap —— UEs are densely distributed around BSs in the central area, while edge BSs only connect to a small number of UEs. This verifies the necessity of optimizing the initial association algorithm described in Section 4.1 \cite{venkateswararao2020traffic,Ren6G}.
\begin{figure}[htb]
	\centering     
	\vspace{0cm}
	\hspace*{0cm}
	\setlength{\abovecaptionskip}{0cm} 
	\setlength{\abovecaptionskip}{0cm} 
	\includegraphics[width=0.5\textwidth]{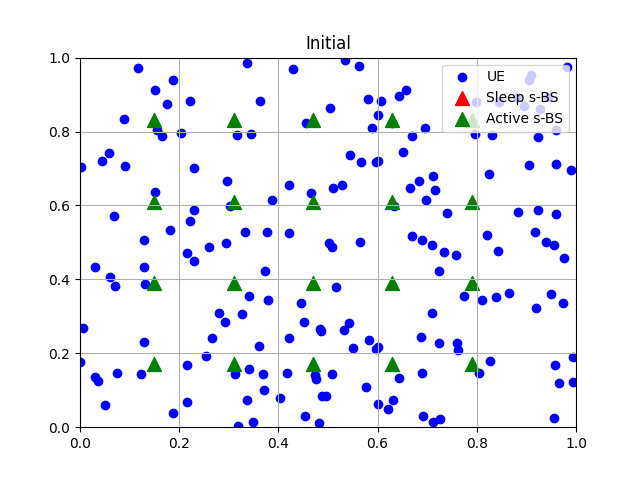}
	\caption{Initial distribution of users and base stations.}
	\label{fig2}
\end{figure}

\begin{figure}[htb]
	\centering     
	\vspace{0cm}
	\hspace*{0cm}
	\setlength{\abovecaptionskip}{0cm} 
	\setlength{\abovecaptionskip}{0cm} 
	\includegraphics[width=0.5\textwidth]{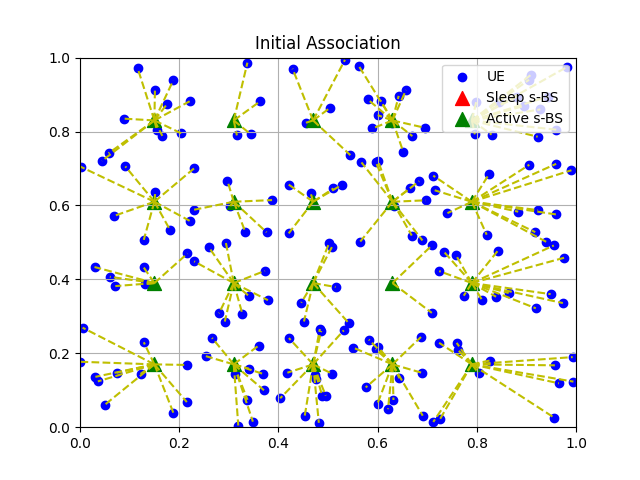}
	\caption{Initial connection of the users to the base stations.}
	\label{fig3}
\end{figure}

\begin{figure}[htb]
	\centering     
	\vspace{-0.2cm}
	\hspace*{0cm}
	\setlength{\abovecaptionskip}{0cm} 
	\setlength{\abovecaptionskip}{0cm} 
	\includegraphics[width=0.5\textwidth]{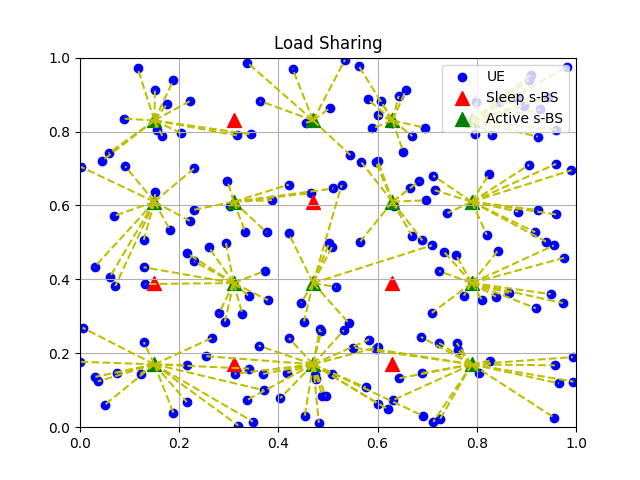}
	\caption{The connectivity of the users to the base stations after performing the base station sleeping decision.}
	\label{fig4}
\end{figure}

\begin{figure}[htb]
	\centering     
	\vspace{0cm}
	\hspace*{0cm}
	\setlength{\abovecaptionskip}{0cm} 
	\setlength{\abovecaptionskip}{0cm} 
	\includegraphics[width=0.5\textwidth]{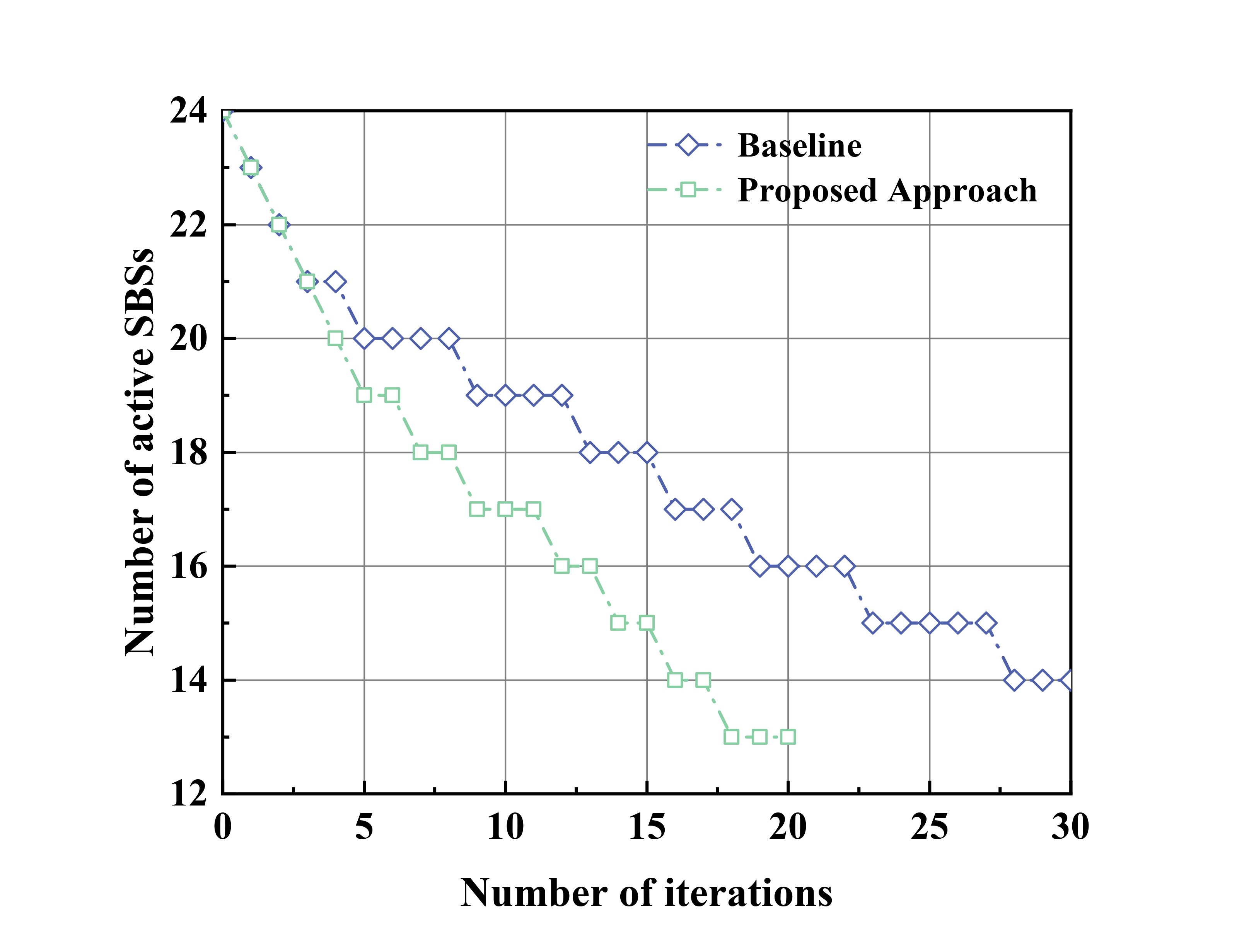}
	\caption{Variation of the number of active base stations during the iterations.}
	\label{fig5}
\end{figure}

\begin{figure}[htb]
	\centering     
	\vspace{0cm}
	\hspace*{0cm}
	\setlength{\abovecaptionskip}{0cm} 
	\setlength{\abovecaptionskip}{0cm} 
	\includegraphics[width=0.5\textwidth]{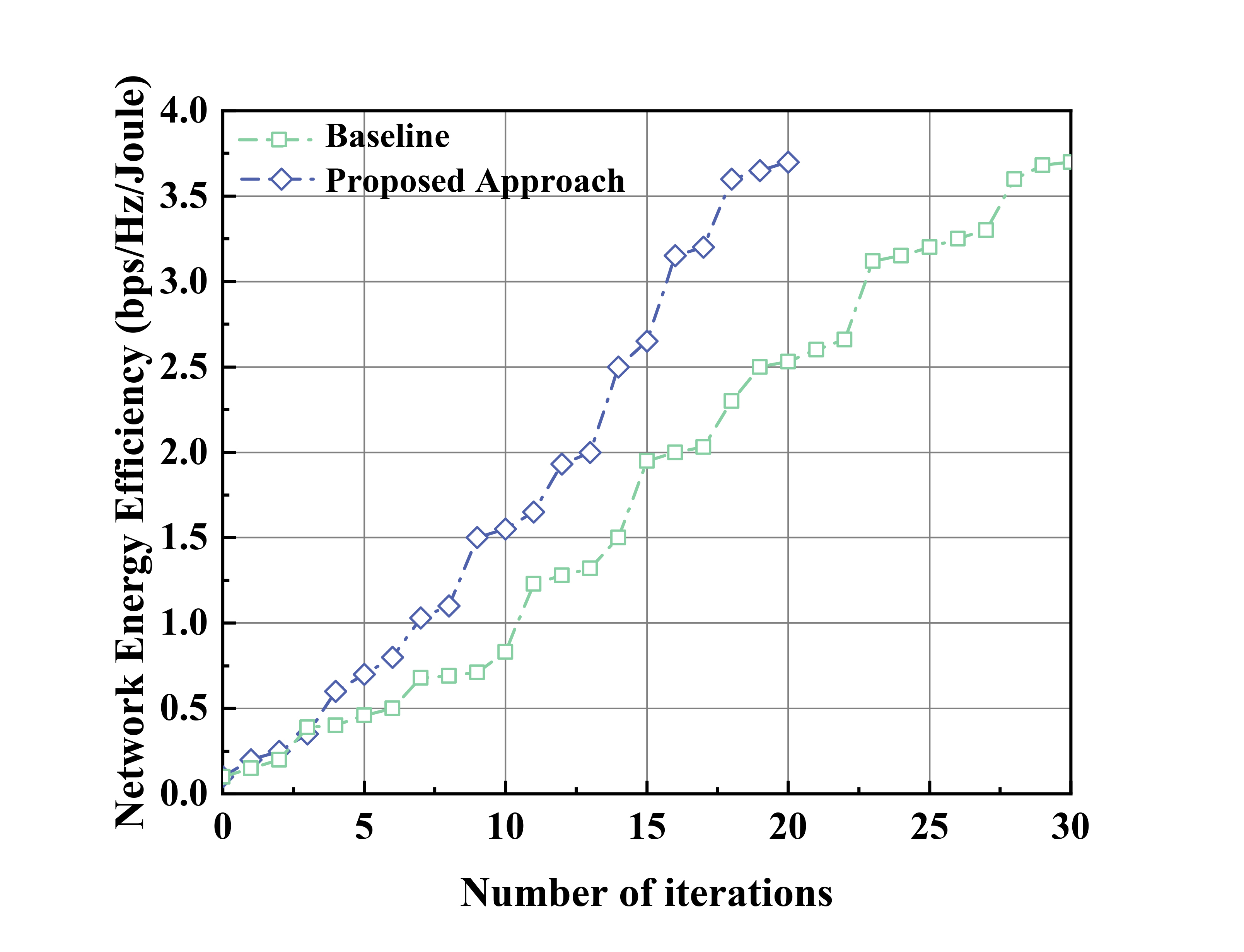}
	\caption{Variation of the energy efficiency during the iterations.}
	\label{fig6}
\end{figure}

Figure 3 presents the optimization results obtained via the UE-BS initial association algorithm (SINR screening and load balancing) described in Section 4.1. The yellow connecting lines clearly illustrate the association relationships between UEs and Active s-BS. At this stage, all UEs are connected to the BSs that offer optimal coverage and relatively low load. No BS is overloaded (with the number of connections $\ge$ $S_{\text{max}} = 30$), and the proportion of low-load BSs (those connecting to $\ge$ 10 UEs) decreases from the initial 60\% to 30\%. This lays a balanced load foundation for the subsequent BS sleeping algorithm.

Fig. 4 illustrates the results after the implementation of the proposed load-sharing-based BS sleeping algorithm: 6 red triangles (representing sleeping small BSs, Sleep s-BS) are successfully deactivated, while the remaining 14 green triangles (Active s-BS) undertake all services. As observed from the topology, the sleeping BSs are concentrated in the edge areas with sparse UE distribution. Additionally, the number of UEs connected to each active BS is evenly distributed between 15 and 22 (corresponding to a load degree of 0.5–0.73), with no cases of load exceeding the limit. This verifies the rationality of the strategy of deactivating low-load BSs and adjacent load sharing.

Figure 5 compares the variation trends of the number of active BSs between the two algorithms during the iteration process. As observed from the figure, the proposed algorithm starts with 20 active BSs, which stably decreases to 13 after approximately 18 iterations, with no subsequent fluctuations in the number of BSs. This reflects the stability of the algorithm brought by its accurate sleeping and load balancing mechanism. The reason lies in that the algorithm accurately locates redundant BSs through load degree calculation (Equation 10) and sleeping index screening (Equation 9). Meanwhile, before load sharing, it conducts a load pre-verification for BSs that take over services, which ensures network stability after BS sleeping. In contrast, the baseline scheme does not stabilize until approximately 28 iterations, with the final number of active BSs being 14. Its convergence process is relatively delayed, primarily because the random sleeping strategy fails to consider load distribution—this leads to unsystematic decision-making in BS sleeping, requiring more iterations to achieve a relatively stable network state. In summary, the proposed scheme exhibits faster convergence speed (converging after about 18 iterations, compared with approximately 28 iterations for the baseline) and results in a smaller number of final active BSs (13 vs. 14). This fully demonstrates the significant advantages of the load-aware BS sleeping strategy in network resource optimization.

Figure 6 presents the variation trend of network EE throughout the iteration process. As indicated by the curve trends, the EE of the proposed algorithm rises significantly faster than that of the baseline algorithm in the early iteration stage (approximately the first 15 iterations) and enters a stable convergence phase earlier. In contrast, the convergence process of the baseline algorithm is relatively delayed. Specifically, relying on the strategy of accurate load assessment and directional load sharing, the proposed algorithm can quickly identify and deactivate redundant BSs while ensuring load balance among the remaining active BSs. This not only reduces the network’s static power consumption but also avoids invalid energy consumption caused by load imbalance, such as service retransmissions and scheduling conflicts—thereby enabling rapid and stable EE improvement in the early iteration stage. By contrast, due to the lack of load-aware sleeping decisions, the baseline algorithm exhibits randomness in selecting BSs for deactivation. This results in a slow EE improvement rate, and its final EE performance after convergence is also significantly lower than that of the proposed algorithm. This result demonstrates that the proposed algorithm has significant advantages in both the convergence speed and final performance of EE. It can achieve network EE optimization more effectively, which holds important practical deployment value for UDN scenarios with strong dynamics.

\section{Summary and Discussion} 
Targeting the issues of high operational costs and low EE caused by the dense deployment of s-BSs in 5G UDNs, the paper proposed an integrated scheme that combined initial connection optimization, with dual constraints of communication quality and load balance, and a load-sharing based BS sleeping strategy, incorporating a novel BS sleeping index and a closed-loop load migration process. A multi-objective mathematical model aiming to maximize EE and minimize the number of active BSs was constructed. Experimental results verified the superiority of the proposed scheme in terms of convergence speed, reduction in the number of active BSs, and EE improvement, thereby providing a feasible path for the green deployment of UDNs.

Future research can be deepened from the following dimensions: First, the current scheme is mainly designed for static scenarios, but in practical UDNs, UE mobility is strong and network topology changes dynamically. In the follow-up, adaptive algorithms integrating real-time UE trajectory tracking and machine learning can be explored to improve the decision-making timeliness of the scheme in dynamic environments. Second, with the penetration of technologies such as the IoT and edge computing, UDNs will face new challenges of massive heterogeneous device access. It is necessary to study how to adapt to the traffic characteristics and service requirements of different devices, and optimize initial connection and sleeping strategies accordingly. Third, the existing models do not fully consider implicit energy consumption such as BS state switching and cross-BS information interaction. In the future, a full-link energy consumption model covering hardware module power consumption and state switching costs can be constructed to enhance the practicality of energy-saving strategies. Fourth, the dense deployment of UDNs intensifies the risks of UE information leakage and network attacks. It is necessary to explore the integration of encryption authentication and privacy protection mechanisms into connection and sleeping processes to ensure the security and compliance of network operations.

\bibliographystyle{ieeetr} 
\bibliography{reference} 
~~~\\
~~~\\

\end{document}